\documentclass[twocolumn,showpacs,preprintnumbers,amsmath,amssymbo]{revtex4}
\usepackage{graphicx}
\usepackage{dcolumn}
\usepackage{bm}
\usepackage{amsmath}
\usepackage{textcomp}
\usepackage{hyperref}
\usepackage{amssymb}
\usepackage{amsfonts}
\usepackage{mathrsfs}
\usepackage[usenames,dvipsnames]{color}

\begin{document}
   \title{Gravitomagnetism and spinor quantum mechanics}
    \author{Ronald J. Adler }
\email{adler@relgyro.stanford.edu}
     \affiliation{1.Gravity Probe B, Hansen Laboratory for Experimental Physics, Stanford University, Stanford CA,
94305\\2.Department of Physics, San Francisco State University, San Francisco CA, 94132}
    \author{Pisin Chen}%
 \email{chen@slac.stanford.edu}
     \affiliation{1. Leung Center for Cosmology and Particle Astrophysics \&
                   Department of Physics and Graduate Institute of Astrophysics, National Taiwan University, Taipei, Taiwan 10617\\
                  2. Kavli Institute for Particle Astrophysics and Cosmology, SLAC National Accelerator Laboratory, Menlo Park CA, 94025}
 \author{Elisa Varani}
\email{elifussin@gmail.com}
\affiliation{Via Ponte Nuovo 24,
29014 Castell'Arquato (PC), Italy}
  %\date{\today}

\begin{abstract}
We give a systematic treatment of a spin $1/2$  particle in a combined electromagnetic field and a weak gravitational field that is produced by a slowly moving matter source. This paper continues previous work on a spin zero particle, but it is largely self-contained and may serve as an introduction to spinors in a Riemann space. The analysis is based on the Dirac equation expressed in generally covariant form and coupled minimally to the electromagnetic field. The restriction to a slowly moving matter source, such as the earth, allows us to describe the gravitational field by a gravitoelectric (Newtonian) potential and a gravitomagnetic (frame-dragging) vector potential, the existence of which has recently been experimentally verified. Our main interest is the coupling of the orbital and spin angular momenta of the particle to the gravitomagnetic field. Specifically we calculate the gravitational gyromagnetic ratio as $\text{ g}_\text{ g}=1$ ; this is to be compared with the electromagnetic gyromagnetic ratio of  $\text{g}_e=2$ for a Dirac electron. 
\end{abstract}

\maketitle

\renewcommand{\thesection}{\arabic{section}}
\renewcommand{\theequation}{\thesection.\arabic{equation}}
\section{Introduction }
Classical systems in external gravitational fields have been studied for centuries, and recently the existence of the gravitomagnetic (or frame-dragging) field caused by the earth's rotation has been observed by the Gravity Probe B (GPB) satellite \cite{Adler2006,WILL1993,Ohanian1976,Everitt2011}. GPB verified the prediction of general relativity for the gravitomagnetic precession of a gyroscope in earth orbit ($42$ mas/yr) to better than $20\%$ \cite{Will43}. Previously, observations of the LAGEOS satellites also indicated the existence of the gravitomagnetic interaction via its effect on the satellite orbits \cite{Will43,Ciufolini1997}. Analysis of the LAGEOS data involves modeling classical effects to very high accuracy in order to extract the gravitomagnetic effect, and the accuracy of the results has been questioned by some authors \cite{Iorio2004}. Analysis of the GPB data also requires highly accurate modeling of classical effects \cite{Will43}. 

While gravitomagnetic effects are generally quite small in the solar system it is widely believed that they may play a large role in jets from active galactic nuclei, so their experimental verification is of more than theoretical interest \cite{Throne2009}.

At the other end of the interest spectrum extensive theoretical work has been done on quantum fields in classical background spaces, the most well known being related to Hawking radiation from black holes \cite{Hawking1970,Birell1982,Adler2001,Adler2006}. However it is important to keep in mind that Hawking radiation has never been observed.  

Interesting experimental work has also been done on quantum systems in the earth's gravitational field, such as neutrons interacting with the earthÕs Newtonian field and atom interferometer experiments aimed at accurately testing the equivalence principle and other subtle general relativistic effects \cite{Nesvizhevdky2002,Dimopoulos2008,Weinberg1972}. There has been some discussion of attempts to see gravitomagnetic effects with these devices but such experiments would be quite difficult due to the small size of the effects and the similarity to classical effects of rotation; this is to be expected since gravitomagnetism manifests itself in a way that is quite similar to rotation, hence the appellation ``frame dragging.'' Laboratory detection of gravitomagnetic effects on a quantum system would clearly be of fundamental interest. 

In this work we give a systematic treatment of a spin $1/2$ particle in a combined electromagnetic field and weak gravitational field; this continues the work of reference \cite{Adler2010}. We describe the particle with the generally covariant Dirac equation in a Riemann space, minimally coupled to the electromagnetic field in the standard gauge invariant way \cite{Bjorken1964, Lawrie1990}. The weak gravitational field is naturally treated according to linearized general relativity theory, and we also assume a slowly moving matter source, such as the earth \cite{Misner1970, Adler1999, Adler2000}. Within this approximation the gravitational field is described by a gravitoelectric (or Newtonian) potential and a gravitomagnetic (or frame-dragging) vector potential, and the field equations are quite analogous to those of classical electromagnetism. We thus refer to it as the gravitoelectromagnetic (GEM) approximation. Our special emphasis throughout this paper is on the gravitomagnetic interaction. 

The paper is organized as follows. After brief review comments on the GEM approximation (section 2) and the Dirac equation in flat space (section 3) we give a detailed discussion of generally covariant spinor theory and the Dirac equation, using the standard approach based on tetrads (sections 4 and 5). We then obtain the limit of the Dirac Lagrangian and the Dirac equation for a weak gravitational field and discuss its interpretation in terms of an energy-momentum tensor (section 6). 

Our discussion of generally covariant spinors and the generally covariant Dirac equation is largely self-contained, and may serve as an introduction to the subject for uninitiated readers. In section 6 we also observe that the non-geometric or ``flat space gravity''  approach of Feynman, Weinberg and others does not appear to be completely equivalent to linearized general relativity theory in its coupling to spin \cite{Feynman1965}. We have not found this discussed in the literature. 

Using the weak gravitational field results we then obtain the non-relativistic limit of the theory (section 7). We do this by integrating the interaction Lagrangian to obtain the interaction energy of the spinor particle with the electromagnetic and the GEM fields, and from that obtain the non-relativistic interaction energies. This allows us to read off, in a simple and intuitive way, the interaction terms that one could use in a non-relativistic Hamiltonian treatment.  In particular we obtain (section 8) the usual anomalous 
g-factor of the electron  $\text{g}_e=2$ and the analogous result for the gravitomagnetic g-factor of a spinor, which is $\text{ g}_\text{ g}=1$.

Section 8 also contains brief comments on the numerical value of some interesting and conceivably observable quantities such as the precession of a spinning particle in the earth's gravitomagnetic field and its relation to the precession of a macroscopic gyroscope; such precession appears to be universal for bodies with angular momentum. The phase shift in an atom interferometer is also  mentioned as an experiment that could, in principle, show the existence of the gravitomagnetic field. 

Lastly it is worth noting what we do {\it not} do in this paper. We study the effect of the gravitational field on a quantum mechanical spinor but not the effect of the spinor on the gravitational field; thus the work does not relate to quantum gravity or quantum spacetime \cite{Oriti2009}. Similarly we do not consider torsion, in which the affine connections have an anti-symmetric part and are not equal to the Christoffel symbols. Torsion does not prove necessary in our discussion, but some authors believe it is necessary in describing the effects of spin on gravity \cite{Trautman2010}. 

\section {The gravitoelectromagnetic (GEM) approximation}

In previous work we discussed linearized general relativity theory for slowly moving matter sources like the earth \cite{Adler2010, Adler1999, Adler2000}. Here we summarize the results very briefly. The metric may be written as the Lorentz metric plus a small perturbation, 
\begin{align}%(2.1)
\text{g}_{\mu\nu}=\eta_{\mu\nu}+h_{\mu\nu}.
\end{align}
We use coordinate freedom to impose the Lorentz gauge condition
\begin{align}%(2.2)
(h_{\mu\nu}-\frac{1}{2}\eta_{\mu\nu}h)^{|\nu}=0,
\end{align}
where the single slash denotes an ordinary derivative. Then the field equations of general relativity tell us that the metric perturbation may be written as 
\begin{align}%(2.3)
h_{\mu\nu}=
\left(
\begin{array}{cccc}
  2\phi& h^1     & h^2      & h^3  \\
  h^1   & 2\phi  & 0           & 0    \\
  h^2   & 0          & 2\phi   & 0  \\
  h^3   & 0          & 0           & 2\phi
\end{array}
\right),\text{\space\space} h_{00}=2\phi,\text{\space\space} h_{0k}=h^k,
\end{align}
where $\phi$  is the Newtonian or gravitoelectric potential  and $h^k \leftrightarrow\vec{h}$ is the gravitomagnetic potential. For slowly moving sources the field equations and the Lorentz condition become\begin{align}%(2.4)
\nabla^2\phi=4\pi G\rho,\text{\space\space}\nabla^2h^j=-16\pi G \rho v^j,\text{\space\space}4\dot{\phi}-\nabla\cdot\vec{h}=0,\text{\space\space}\dot{\vec{h}}=0,
\end{align}				
 where $\rho$ is the source mass-energy density and  $v^j$  is its velocity. 

The physical fields, which exert forces on particles, are the gravitoelectric (or Newtonian) field and the gravitomagnetic (or frame-dragging) field, which are defined by
\begin{align}%(2.5)
\vec{\text{g}}=\nabla\phi,\text{\space\space}\vec{\Omega}=\nabla \times\vec{h}.
\end{align}	
We call this equation system the gravitoelectromagnetic or GEM limit because of its close similarity to classical electromagnetism.

\section{Flat space Dirac equation and the non-relativistic limit}
\setcounter{equation}{0}
In this section we discuss the Dirac equation in the flat space of special relativity and recast it into a Schroedinger equation form (SEF), which provides one convenient way to obtain the non-relativistic limit \cite{Bjorken1964}. The SEF is exact and involves only the upper two components of the spinor wave function - the relevant components for positive energy solutions in the non-relativistic limit. One reason for doing this is to serve as a basis of comparison for the alternative method we will use in section 6 when we discuss gravitational interactions. Throughout this section  $\gamma^\mu$ denotes the flat space Dirac matrices \cite{Peskin1995, Bjorken1964}. 

The Dirac Lagrangian and the Euler-Lagrange equations that follow from it are 
\begin{subequations}
\begin{align}%(3.1)
&L=a\bar{\psi}(i\gamma^\mu\vec{\partial_\mu}-m)\psi+b\bar{\psi}(-i\gamma^\mu\overleftarrow{\partial}_\mu-m)\psi-eA_\mu\bar{\psi}\gamma^\mu\psi,\\
&(i\gamma^\mu\partial_\mu-m)\psi=eA_\mu\gamma^\mu\psi,\text{\space\space}\bar{\psi}(-i\gamma^\mu\overleftarrow{\partial}_\mu m)=eA_\mu\bar{\psi}\gamma^\mu.
\end{align}
The spinor and its adjoint are considered independent in obtaining (3.1b). The constants  $a$  and  $b$ are arbitrary, so long as $a+b\neq0$ . The $\gamma^\mu$  obey the flat space Dirac algebra, 
\begin{align}
\{\gamma^\mu,\gamma^\alpha\}=2\eta^{\mu\nu}I.
\end{align}
\end{subequations}
The adjoint spinor is assumed to be related to the spinor by a linear metric relation, $\bar{\psi}=\psi^\dagger M$  where M  is to be determined; consistency of the equations (3.1b) is then assured if M obeys
\begin{align}%(3.2)
M^{-1}\hat{\gamma}^{\mu^\dagger}M=\hat{\gamma}^\mu,\text{\space\space}M^{-1}=M=\hat{\gamma}^0,\text{\space\space}\bar{\psi}=\psi^\dagger\gamma^0.
\end{align}
Eq. (3.2) is easy to verify for the choice of gamma matrices given below in (3.4).

The Hamiltonian form of the Dirac equation will be useful for studying interaction energies in this section. It is gotten by multiplying (3.1) by  $\gamma^0$  to obtain
\begin{align}%(3.3)
i\partial_t\psi=\beta m\psi+V+\vec{\alpha}\cdot\vec{\Pi}\psi,\text{\space\space}\beta\equiv\gamma^0,\text{\space\space}\alpha\equiv\gamma^0\gamma^k,\text{\space\space}\vec{p}\equiv-i\nabla.
\end{align}
Pauli's choice of gamma matrices is natural for our later discussion of the non-relativistic limit,  	
\begin{align}%(3.4)
\beta=\gamma^0=\left(
\begin{array}{cc}
  I         & 0   \\
  0        & I  
\end{array}
\right),
\text{\space\space}\gamma^i=\left(
\begin{array}{cc}
  0  & \sigma^i     \\
  -\sigma^i   & 0          
\end{array}
\right),\text{\space\space}\vec{\alpha}\equiv\left(
\begin{array}{cc}
  0              & \sigma     \\
  \sigma   & 0  
\end{array}
\right).
\end{align}

Next we break the 4-component wave function $\psi$  into two
2-component Pauli spinor wave functions and also factor out the time dependence due to the rest mass   by substituting 
 \begin{align}%(3.5)
\psi=e^{-imt}\left(
\begin{array}{c}
  \Psi   \\
  \varphi          
\end{array}
\right),
\end{align}
which leads to the coupled equations, 
\begin{align}%(3.6)
i\partial_t\Psi=V\Psi+(\vec{\sigma}\cdot\vec{\Pi})\varphi,\text{\space\space}i\partial_t\phi+2m\varphi-V\varphi=(\vec{\sigma}\cdot\vec{Pi})\Psi.
\end{align}
We are interested in  $\Psi$  so we solve for $\varphi$ , and obtain symbolically,
\begin{subequations}
\begin{align}%(3.7)
i\partial_t\Psi=V\Psi+(\vec{\sigma}\cdot\vec{\Pi})(2m-V+i\partial_t)^{-1}(\vec{\sigma}\cdot\vec{\Pi})\Psi,\\
\varphi=(2m-V+i\partial_t)^{-1}(\vec{\sigma}\cdot\vec{\Pi})\Psi.
\end{align}
\end{subequations}
The inverse operator $(2m-V+i\partial_t)^{-1}$  may be defined by its expansion in the time derivative, as discussed in Appendix A. Note that Eq. (3.7a) is an exact equation for $\Psi$, although it is of infinite order in the time derivative. 

For the special case of a free particle the operator factors on the right side of (3.7a) commute and it becomes simply
\begin{align}%(3.8)
i\partial_t\psi=(i\partial_t+2m)^{-1}\vec{p}^2\psi.
\end{align}
However the operators on the right side of (3.7a) will not in general commute unless the field $A_{\mu}$  is constant.

In a low velocity system the time variations of $\Psi$  and $V$ are associated with non-relativistic energies, which are much less than the rest energy $m$, so we may approximate (3.7a) by
\begin{align}%(3.9)
i\partial_t\Psi=V\psi+\frac{(\vec{\sigma}\cdot\vec{\Pi})^2}{2m}\Psi.
\end{align}
This is the Schroedinger equation for spin $1/2$ particles, often called the Pauli equation. The Pauli equation shows clearly how the spin and orbital angular momentum interact with the magnetic field. Pauli spin matrix algebra leads to an illuminating form for (3.9): to lowest order in ${e}$, 
\begin{align}%(3.10)
i\partial_t\Psi&=V\Psi+\frac{\vec{\Pi}^2}{2m}\Psi-\frac{e\vec{B}\cdot\sigma}{2m}\Psi\notag\\
&=V\Psi+\frac{\vec{p}^2}{2m}\Psi-\frac{e\vec{A}\cdot\vec{p}}{m}\Psi-\frac{e\vec{B}\cdot\sigma}{2m}\Psi,
\end{align}
where we have used the Lorentz gauge in which  $\nabla\cdot\vec{A}=-\dot{A}^0$  and assumed the Coulombic $A^0$  has negligible time dependence. The gyromagnetic ratio or g-factor of a particle or system is defined in terms of its magnetic moment $\vec{\mu}$  and angular momentum $\vec{J}$  by $\vec{\mu}=\text{g}_e(e/2m)\vec{J}$
 ; thus, from (3.10), the fact that the energy is $-\vec{\mu}\cdot\vec{B}$, and the electron spin of $\vec{S}=\sigma/2$  it is evident that the electron g-factor is g$_e=2$. 
 
The relative coupling of the spin and orbital magnetic moments is made most clear if we consider a magnetic field that is approximately constant over the size of the system, in which case we can choose  $\vec{A}=(\vec{B}\times\vec{r})/2$ and find from (3.10)
\begin{align}%(3.11)
i\partial_t\Psi&=V\Psi+\frac{\vec{p}^2}{2m}\Psi-\frac{e\vec{B}}{2m}(2\vec{S}+\vec{L})\Psi,\notag\\
\vec{S}&=\sigma/2,\text{\space\space}\vec{L}=\vec{r}\times\vec{p}.
\end{align}
That is g$_e=2$  for the electron spin and g$_e=1$  for the orbital angular momentum. 

Equation (3.7a) may be expanded to higher order to study such things as hyperfine structure and relativistic corrections in the hydrogen atom spectrum \cite{Shankar1994}. That is 
\begin{align}%(3.12)
i\partial_t\Psi=V\Psi+\frac{(\vec{\sigma}\cdot\vec{\Pi})^2}{2m}\Psi-\frac{(\vec{\sigma}\cdot\vec{\Pi})(i\partial_t-V)(\vec{\sigma}\cdot\vec{\Pi})}{4m^2}\Psi,
\end{align}
However an important problem and caveat is that the wave function  $\Psi$ in (3.12) is only the upper half of the Dirac wave function, so the quantity that must be normalized is $|\Psi|^2+|\varphi|^2$  rather than $|\psi_s|^2$  for a Schroedinger or Pauli wave function $\psi_s$  . Thus to insure Hermiticity and conserve probability one must renormalize the wave function as discussed in detail in ref. \cite{Shankar1994}. It is for this reason that we will adopt an alternative and conceptually simpler approach to the non-relativistic limit in section 7. 

\section{Generally covariant spinor theory}
\setcounter{equation}{0}
The gravitational interaction of a spinor may be obtained most easily by making the Dirac Lagrangian (3.1a) and Dirac equation (3.1b) generally covariant. To do this we adopt the standard approach of using a tetrad of basis vectors in order to relate the generally covariant theory to the special relativistic theory in Lorentz coordinates \cite{Weinberg1972, Lawrie1990}. This is a most natural, almost inevitable, approach since Dirac spinors transform by the lowest dimensional representation  $S$ of the Lorentz group; that is $\psi'=S\psi$ .  

Two properties of the Dirac Lagrangian and Dirac equation must be modified to obtain a generally covariant theory: the Dirac algebra in (3.2) must be made covariant and the derivative of the spinor in (3.1) must be made into a covariant derivative. We will discuss both in detail. 

The Dirac algebra (3.1c) is easily made covariant by replacing the Lorentz metric $\eta^{\mu\nu}$  by the Riemannian metric g$_{\mu\nu}$ ,  
\begin{align}%(4.1)
\{\gamma^\mu,\gamma^{\alpha}\}=2\text{g}^{\mu\nu}I.
\end{align}
A set of $\gamma^\mu$  matrices that satisfy (4.1) is easily constructed by using a set of constant  $\hat{\gamma}^b$ that satisfies the special relativistic relation (3.2) and a tetrad field  $e^\mu_b$  normalized by the usual tetrad relations
\begin{align}%(4.2)
e^\mu_b e^\nu_\text{a}\text{g}_{\mu\nu}=\eta_{\text{a}b},\text{\space\space}\text{g}^{\alpha\beta}=e^{\alpha}_ce^{\beta}_d\eta^{cd}.
\end{align}
Here the Greek indices label components of the tetrad vectors and Latin indices label the vectors. In terms of a convenient set of constant Dirac matrices $\hat{\gamma}^b$, such as those in (3.4), we define the   $\gamma^\mu$ by 
\begin{align}%(4.3)
\gamma^\mu=e^\mu_b\hat{\gamma}^b.
\end{align}
It then follows from (3.1c) and (4.2) that the $\gamma^\mu$  satisfy 
\begin{align}%(4.4)
\{\gamma^\mu,\gamma^\nu\}=e^\mu_be^\nu_\text{a}\{\hat{\gamma}^b,\hat{\gamma}^\text{a}\}=e^\mu_be^\nu_\text{a}2\eta^{\text{a}b}I=2\text{g}^{\mu\nu}I.
\end{align}

The covariant derivative of a spinor is defined so as to transform as a vector under general coordinate transformations and as a spinor under Lorentz transformation of the tetrad basis.  As with the covariant derivative of a vector we define a rule for transplanting a spinor from $x$ to a nearby point $x+dx$ , 
\begin{align}%(4.5)
\psi^*(x+dx)=\psi(x)-\Gamma_\mu\psi(x)dx^\mu.
\end{align}
The matrices $\Gamma_a$ are variously called spin connections, affine spin connections, or Fock-Ivanenko coefficients. The covariant derivative is then defined in terms of the difference between the value of the spinor and the value it would have if transplanted to the nearby point. That is 
\begin{align}%(4.6)
\psi(x)_{||\nu}dx^\nu&=[\psi(x)+\psi(x)_{|\nu}dx^\nu]-[\psi(x)-\Gamma_\nu(x)\psi(x)dx^\nu]\notag\\
                                          &=[\psi(x)_{|\nu}+\Gamma_\nu(x)\psi(x)]dx^\nu,\notag\\
\psi_{||\nu}&=\psi_{|\nu}+\Gamma_\nu\psi=(\partial_\nu+\Gamma_\nu)\psi\equiv D_\nu\psi.                                       
\end{align}
Here the double slash denotes a covariant derivative. Since the spinor covariant derivative must transform as a vector under coordinate transformations and as a spinor under Lorentz transformations of the tetrad basis, we have 
\begin{align}%(4.7)
\psi'_{||\mu}=\frac{\partial x^\nu}{\partial x'^\mu}S\psi_{||\nu},                                    
\end{align}
It follows from (4.6) and (4.7) that the spin connections must transform according to 
\begin{align}%(4.8)
\Gamma'_\nu=\frac{\partial x^\nu}{\partial x'^\mu}[S\Gamma_\nu S^{-1}-S_{|\nu}S^{-1}].
\end{align}
The transformation (4.8) is formally similar to that of the affine connections used for vector covariant derivatives. 

The covariant derivative of an adjoint spinor follows easily from that of a spinor in (4.6); we ask that the inner product $\bar{\psi}\chi$ of a spinor $\chi$  and an adjoint spinor $\bar{\psi}$  be a scalar and thus have a covariant derivative $(\bar{\psi}\chi)_{||\mu}$ equal to the ordinary derivative $(\bar{\psi}\chi)_{|\mu}$, and we also ask that the product rule hold for both the ordinary and the covariant derivatives. The result is   
\begin{align}%(4.9)
\bar{\psi}_{||\mu}=\bar{\psi}_{|\mu}-\bar{\psi}\Gamma_\mu.
\end{align}
The same idea leads to the covariant derivative of a gamma matrix, with only a bit more algebra; that is we ask that the expression $(\bar{\psi}\gamma^\mu\chi)_{||\alpha}$ be a second rank tensor and that it obey the product rule of differentiation, and find from (4.6) and (4.9)
\begin{align}%(4.10)
\gamma^\mu_{||\omega}=\gamma^\mu_{|\omega}+\left\{
\begin{array}{c}
\mu\\
\omega\sigma
\end{array}
\right\}+[\Gamma_\omega,\gamma^\mu].
\end{align}
This expression plays an important role in obtaining the spin connections in the next section.

\section{Covariant Dirac Lagrangian and Dirac equation }
\setcounter{equation}{0}

In this section we give a covariant Lagrangian and obtain the covariant Dirac equation. In the process we get a relation between the spinor and its adjoint (i.e. a spin metric) and evaluate the spin connections.  

The choice of a covariant Dirac Lagrangian $L$, and its associated Lagrangian density $\mathcal{L}$, is rather obvious from the flat space Lagrangian in (3.1), 
\begin{align}%(5.1)
L=a\bar{\psi}(i\gamma^\mu\psi_{||\mu}-m\psi)+b(-i\bar{\psi}_{||\mu}\gamma^\mu-\bar{\psi}m)\psi,\text{\space\space}\mathcal{L}=\sqrt{g}L.
\end{align}
Coupling to the electromagnetic field will be included later. The $\gamma^\mu$ denotes the {\it covariant} Dirac matrices (4.3) throughout this section. The Dirac equations for the spinor and the adjoint spinor follow directly as the Euler-Lagrange equations of the Lagrangian density $\mathcal{L}$ with $\psi$  and $\bar{\psi}$ treated as independent variables, 
\begin{subequations}
\begin{align}%(5.2)
(a+b)(i\gamma^\mu\psi_{||\mu}-m\psi)+ib\gamma^\mu_{\text{\space}||\mu}\psi=0\\
(a+b)(\bar{\psi}_{||\mu}i\gamma^\mu+m\psi)+ia\bar{\psi}\gamma^\mu_{\text{\space}||\mu}=0.
\end{align}
\end{subequations}
For simplicity we assume that the spin connections, unspecified up to this point, may be chosen so that the divergence of $\gamma^\alpha$  that appears in (5.2) vanishes, $\gamma^\mu_{\text{\space}||\mu}=0$. The covariant Dirac equation is then the obvious generalization of the flat space equations (3.1). The spin connections will be obtained below. Also for simplicity and symmetry we choose henceforth $a=b=1/2$; this will prove convenient later. 

Next, as in flat space in section 3, we ask that there be a relation between the adjoint and the spinor, $\bar{\psi}=\psi^\dagger M$, such that the two equations (5.2) are consistent. Manipulating (5.2a) we get for the adjoint, 
\begin{align}%(5.3)
-i\bar{\psi}_{|\mu}\tilde{\gamma}^\mu-i\bar{\psi}M^{-1}_{\text{\space}|\mu}M\tilde{\gamma}^\mu-i\bar{\psi}\tilde{\Gamma}_\mu\tilde{\gamma}^\mu-\bar{\psi}m=0,\notag\\
\tilde{\gamma}^\mu\equiv M^{-1}\gamma^{\mu^\dagger}M,\text{\space\space}\tilde{\Gamma}_\mu\equiv M^{-1}\Gamma_\mu^{\text{\space}\dagger}M.
\end{align}
We then compare (5.3) with (5.2b), written as 
\begin{align}%(5.4)
-i\bar{\psi}_{|\mu}\gamma^\mu+i\bar{\psi}\Gamma_\mu\gamma^\mu-\bar{\psi}m=0,
\end{align}
and see that $M$ must satisfy the following two equations 
\begin{subequations}
\begin{align}%(5.5)
\gamma^\mu=\tilde{\gamma}^\mu=M^{-1}\gamma^{\mu^\dagger}M,\\
-\Gamma_\mu=\tilde{\Gamma}_\mu=M^{-1}\Gamma_\mu^{\text{\space\space}\dagger}M+M^{-1}_{\text{\space\space\space}|\mu}M.
\end{align}
\end{subequations}
Eq. (5.5a) may be written in terms of flat space $\hat{\gamma}^b$ as 
\begin{align}%(5.6)
e^\mu_b\hat{\gamma}^b=e^\mu_bM^{-1}\hat{\gamma}^{b^\dagger}M.
\end{align}
Thus it is obvious that we should ask $\hat{\gamma}^b=M^{-1}\hat{\gamma}^{b^\dagger}M$, which is the same as in the case of flat space and special relativity (3.2), so we may choose $M^{-1}=M=\hat{\gamma}^0$. Then the derivative of $M$ is zero, and it is easy to verify that the choice $M^{-1}=M=\hat{\gamma}^0$ also satisfies (5.5b). 

Our remaining task is to obtain specific spin connections $\Gamma_\alpha$. To do this we make the natural demand that $\gamma^\mu$ have a null covariant derivative, so from (4.10)
\begin{align}%(5.7)
\gamma^\mu_{||\alpha}=\gamma^\mu_{|\alpha}+\left\{
\begin{array}{c}
\mu\\ \alpha\beta
\end{array}
\right\}\gamma^\beta+[\Gamma_\alpha,\gamma^\mu]=0.
\end{align}
This guarantees that the divergence vanishes, $\gamma^\mu_{\text{\space}||\mu}=0$, as we have already mentioned.  However it is a stronger demand analogous to the standard demand in general relativity that the metric have a null covariant derivative, which forces the affine connections to be the Christoffel symbols. Note also that $\Gamma_\alpha$  is obviously arbitrary up to a multiple of the identity, which we will suppress henceforth. 

To solve (5.7) we express $\gamma^\mu$ in terms of flat space gammas $\hat{\gamma}^b$ as in (4.3) and rewrite (5.7) as
\begin{align}%(5.8)
e^\mu_{b||\alpha}\hat{\gamma}^b+[\Gamma_\alpha,\hat{\gamma}^b]e^\mu_b=0.
\end{align}
Multiplying this by the inverse tetrad matrix we get 
\begin{align}%(5.9)
[\Gamma_\alpha,\hat{\gamma}^c]=-e^c_\mu e^\mu_{b||\alpha}\hat{\gamma}^b.
\end{align}
We next note the well-known commutation relation on the sigma matrices, which are defined as $\hat{\sigma}^{\text{a}b}\equiv(i/2)[\hat{\gamma}^\text{a},\hat{\gamma}^b]$,
\begin{align}%(5.10)
[\hat{\sigma}^{\text{a}b},\hat{\gamma}^c]=2i(\hat{\gamma}^\text{a}\eta^{bc}-\hat{\gamma}^b\eta^{\text{a}c}),
\end{align}
From (5.10) it is evident that we should seek a solution that is proportional to $\hat{\sigma}^{\text{a}b}$ times a product of the tetrad and its derivatives. It is easy to verify that the specific choice 
\begin{align}%(5.11)
\Gamma_\alpha=\frac{i}{4}e_{b\mu}e^\mu_{\text{a}||\alpha}\hat{\sigma}^{\text{a}b},
\end{align}
satisfies (5.9) and thus serves as the spin connection. 

We thus have obtained a generally covariant theory in which the Lagrangian, the Dirac equations, the relation of the spinor to its adjoint, and the spin connections are generally covariant and consistent. 

Finally we include coupling to the electromagnetic field in the usual minimal coupling way, that is by substituting $iD_\mu \rightarrow iD_\mu-eA_\mu$; this gives the complete covariant Lagrangian
\begin{align}%(5.12)
L=\frac{1}{2}\bar{\psi}(i\gamma^\mu\psi_{||\mu}-m\psi)+\frac{1}{2}(-i\bar{\psi}_{||\mu}\gamma^\mu-\bar{\psi}m)\psi-eA_\mu\bar{\psi}\gamma^\mu\psi.
\end{align}
We will study the weak gravitational field limit of this in the next section. 

\section{Linearized theory for weak gravity}
\setcounter{equation}{0}
In this section we use the results of section 5 for covariant spinor theory to work out the weak field linearized theory. This is done by setting up an appropriate tetrad and using it to expand the Lagrangian (5.12) to lowest order in the metric perturbation. The result is that there are three interaction terms in the Lagrangian, one associated with the spin coefficients and the second with the alteration in the $\gamma^\mu$ caused by gravity. Remarkably the first vanishes in the linearized theory, while the second corresponds to an interaction via the energy momentum tensor, as intuition should suggest. The third term is a cross term between the weak gravity and electromagnetic fields. 

In a space with a nearly Lorentz metric (2.1) it is natural to choose a tetrad that lies nearly along the coordinate axes, 
\begin{align}%(6.1)
e^\mu_a=\delta^\mu_a+w^\mu_a,\text{\space\space}e^b_\nu=\delta^b_\nu-w^b_\nu,
\end{align}
where $w^\mu_a$ is a small quantity to be determined. From the fundamental tetrad relation (4.2) it follows that we should choose a symmetric $w_{\mu\nu}=-(1/2)h_{\mu\nu}$ and thus have a tetrad and $\gamma^\mu$ matrices given by
\begin{align}%(6.2)
e^\mu_a&=\delta^\mu_a-(1/2)h^\mu_a,\notag\\
\gamma^\mu&=[\delta^\mu_a-(1/2)h^\mu_a]\hat{\gamma}^a=\hat{\gamma}^\mu-(1/2)h^\mu_a\hat{\gamma}^a.
\end{align}
Since Greek tensor indices and Latin tetrad indices are intimately mixed in the linearized theory we will not distinguish between them in this section. 

To evaluate the spin connections (5.11) with the tetrad (6.2) we need the Christoffel symbols and the covariant derivative of the tetrad to first order in  $h_{\mu\nu}$, 
\begin{align}%(6.3)
\left\{
\begin{array}{c}
\nu\\ \mu\omega
\end{array}
\right\}=(1/2)(h_{\omega\text{\space}|\mu}^{\text{\space}\nu}+h_{\mu\text{\space}|\omega}^{\text{\space}\nu}-h_{\mu\omega}^{\text{\space\space\space}|\nu}),\notag\\
e^\nu_{\text{a}||\mu}=(1/2)(h_{\mu\text{\space}|\text{a}}^{\text{\space}\nu}-h_{\mu\text{a}}^{\text{\space\space\space}|\nu}).
\end{align}
From (5.11), (6.2) and (6.3) we obtain the spin connections, 
\begin{align}%(6.4)
\Gamma_\mu=\frac{i}{4}e_{b\nu}e^\nu_{\text{a}||\mu}\hat{\sigma}^{\text{a}b}\cong\frac{i}{4}h_{\mu b|\text{a}}\hat{\sigma}^{\text{a}b}.
\end{align}
Thus the Dirac Lagrangian (5.12) becomes, 
\begin{align}%(6.5)
L=\frac{1}{2}\bar{\psi}(i\gamma^\mu\psi_{|\mu}-m\psi)+\frac{1}{2}(-i\bar{\psi}_{|\mu}\gamma^\mu-\bar{\psi}m)\psi-eA_\mu\bar{\psi}\gamma^\mu\psi\notag\\
+\frac{i}{2}\bar{\psi}\{\hat{\gamma},\Gamma_\mu\}\psi-\frac{i}{4}h^\mu_{\text{\space}\alpha}[\bar{\psi}\hat{\gamma}^\alpha\psi_{|\mu}-\bar{\psi}_{|\mu}\hat{\gamma}^\alpha\psi]+\frac{1}{2}h^\mu_{\text{\space}\alpha}A_\mu\bar{\psi}\hat{\gamma}^\alpha\psi,
\end{align}
with $\Gamma_\mu$ given in (6.4). The first line is the Dirac Lagrangian in flat space (3.1a), and the other three terms are gravitational interactions that we now address. 

The first interaction term in the second line of (6.5), due to the spin connections, contains the anti-commutator $\{\hat{\gamma}^\mu,\Gamma_\mu\}$. With the use of the symmetry of $h_{\mu\nu}$, the Dirac algebra (3.1c), and the operator identity $[AB,C]=A\{B,C\}-\{A,C\}B$ it is straightforward to verify the following two expressions, 
\begin{align}%(6.6)
h_{\mu b|a}\hat{\gamma}^\mu\hat{\sigma}^{ab}=i(h^a_{\text{\space}b|a}-h_{|b})\hat{\gamma}^b,\notag\\
h_{\mu b|a}\hat{\sigma}^{ab}\hat{\gamma}^\mu=i(h_{|b}-h^a_{\text{\space} b|a})\hat{\gamma}^b,
\end{align}
and thereby see that
\begin{align}%(6.7)
\{\hat{\gamma}^\mu,\Gamma_\mu\}=\frac{i}{4}h_{\mu b|a}\{\hat{\gamma}^\mu,\hat{\sigma}^{ab}\}=0.
\end{align}
Thus the interaction term containing the spin connections in (6.5) vanishes, which is a remarkable simplification. It should be stressed that this is only true to first order, and the spin connections will generally be of interest in the full theory.

There remains in the Lagrangian (6.5) only interactions due to the modification of the $\hat{\gamma}^\mu$ by gravity in (6.2); $L$ may now be written as 	
\begin{align}%(6.8)
L=\frac{1}{2}\bar{\psi}(i\hat{\gamma}\psi_{|\mu}-m\psi)+\frac{1}{2}(-i\bar{\psi}_{|\mu}\hat{\gamma}-\bar{\psi}m)\psi-eA_\mu\bar{\psi}\hat{\gamma}^\mu\psi\notag\\
-\frac{1}{2}h^\mu_{\text{\space}\alpha}[\frac{1}{2}\bar{\psi}\hat{\gamma}^\alpha(i\psi_{|\mu}-eA_\mu\psi)-\frac{1}{2}(i\bar{\psi}_{|\mu}+eA_\mu\bar{\psi})\hat{\gamma}^\alpha\psi]
\end{align}	
The quantity in brackets in (6.8) is the appropriately symmetrized energy-momentum tensor $T^a_{\text{\space}\mu}$ for the Dirac field interacting with the electromagnetic field; that is, the gravitational interaction Lagrangian may be expressed as 
\begin{align}%(6.9)
L_{IG}=-\frac{1}{2}h^\mu_{\text{\space}\alpha}[\frac{1}{2}\bar{\psi}\hat{\gamma}^\alpha(i\psi_{|\mu}-eA_\mu\psi)-\frac{1}{2}(i\bar{\psi}_{|\mu}+eA_\mu\bar{\psi})\hat{\gamma}^\alpha\psi]\notag\\
=-\frac{1}{2}h_{\mu\alpha}T^{\mu\alpha}.
\end{align}	
The energy momentum tensor is discussed further in Appendix B. 	
	
The interaction (6.9) consists of the inner product of the field $h_{\mu\nu}$  with the conserved energy-momentum tensor $T^{\mu\nu}$; this coupling is in close analogy with the electromagnetic coupling between the field $A_\mu$ and the conserved current $j^\mu=e\bar{\psi}\gamma^\mu\psi$ in (6.8). Feynman has emphasized this analogy and developed a complete ``flat space'' gravitational theory, with gravity treated as an ``ordinary'' two index (spin 2) field and formulated by analogy with electromagnetism, at least to lowest order \cite{Feynman1965}. The geometric interpretation of gravity is thereby suppressed or ignored. Weinberg has similarly stressed that the geometric interpretation of gravity is not essential \cite{Weinberg1972, Feynman1965}. Schwinger also has used a similar and probably equivalent non-geometric methodology called source theory to obtain the standard results of general relativity theory, including the precession of a gyroscope due to the gravitomagnetic field \cite{Schwinger1976}. However there is a problem with relating the geometric and non-geometric viewpoints, in that the Euler-Lagrange field equations are based on the Lagrangian $\it{density}$ $\mathcal{L}\sqrt{\text{g}}L\cong(1+h/2)L$ and not the Lagrangian $L$, so there is an additional interaction term $(h/2)L$ in the geometric theory that is not present in the non-geometric theory; the equivalence of the Feynman approach to the linearized geometric approach is thus spoiled whenever the additional term does not vanish. 	
	
The difference between the Dirac equation per our geometric development and that which one would obtain from the non-geometric approach is easy to see. The Dirac equation that follows from (6.8) with $\mathcal{L}=\sqrt{\text{g}}L\cong(1+h/2)L$ is 	
\begin{align}%(6.10)
\gamma^\mu(i\psi_{|\mu}&-eA_\mu)-m\psi\notag\\
&=\frac{1}{2}h_{\mu\nu}\hat{\gamma}^\mu(i\psi^{|\nu}-eA^\nu\psi)+\frac{1}{4}(h^\mu_{\text{\space}\nu|\mu}-h_{|\nu})i\hat{\gamma}^\nu\psi.
\end{align}	
The last term on the right containing $h_{|\nu}$ would not be present in the non-geometric approach. This will be discussed further in section 7. 	
	
In summary of this section, the Lagrangian (6.8) contains the interaction of the Dirac field with the electromagnetic field to all orders and the interaction with the gravitational field only to lowest order; (6.10) is the corresponding Dirac equation. We will discuss the interaction energies further in the following section in which we consider the non-relativistic or low velocity limit of the theory. 	

\section{Non-relativistic limit}
\setcounter{equation}{0}	
	
We wish to use the results of the previous sections to obtain a non-relativistic limit of the theory and calculate in a simple way some interesting properties of a spin $1/2$ particle such as the electromagnetic g-factor and its gravitomagnetic analogue. The most familiar approach to this problem is to work with the upper two components of the Dirac wave function as we did in section 3, and take the non-relativistic limit \cite{Bjorken1964, Shankar1994}. However the alternative approach we use in this section is conceptually simpler and avoids the problems of renormalization and Hermiticity that occur in the approach of sec. 3. The basic idea is to integrate the interaction Lagrangian over 3-space to get the interaction energy, then put the energy expression with Dirac 4-spinor wave functions, into a form using Pauli 2-spinor wave functions, all in the low velocity limit. 	
	
In this section we will always work in nearly flat space with Lorentz coordinates; the Dirac $\gamma^\mu$ will be those of flat space and no hat will be used. Moreover for simplicity we will work in the Lorentz gauge for both the electromagnetic and GEM fields, and take both the Coulomb potential $A^0$ and the Newtonian potential $\phi$  to have negligible time dependence; that is $\dot{A}^0=-\nabla\cdot\vec{A}=0$ and $4\dot{\phi}=\nabla\cdot\vec{h}=0$. This is quite appropriate, for example, for electromagnetic interactions in atoms and GEM interactions on the earth. 	
	
To illustrate the method we first consider only the electromagnetic interaction in flat space; the results will be the same as those in section 3, in particular $\text{g}_e=2$. The interaction Lagrangian and the interaction energy are, from (6.8),   	
\begin{subequations}
\begin{align}%(7.1)
L_{IEM}=-eA_\mu(\bar{\psi}\gamma^\mu\psi)=-A_\mu j^\mu,\\
\Delta E_{EM}=-\int L_{IEM}d^3x.
\end{align}	
\end{subequations}	
For the Dirac $\psi$ we use a convenient device, an expansion in terms of free positive energy Dirac wave functions on the mass shell. That is 	
\begin{align}%(7.2)
\psi=\sum_{s=1,2}\limits\int\frac{d^3p}{(2\pi)^3}f(p,s)[e^{ip_\alpha}x^\alpha u(p,s)],\notag\\
E^2=(p^0)^2=p^2+m^2.
\end{align}		
The positive energy wave functions do not form a complete set, but the approximation (7.2) should be quite good for distances much larger than the Compton wavelength, $\hbar/m$; (7.2) is our fundamental assumption. A key idea in the calculation is to express the Dirac 4-spinor $u(p,s)$ in terms of a Pauli 2-spinor $\chi_s$ \cite{Adler1966},   	
\begin{align}%(7.3)
e^{-ip_\alpha x^\alpha}u(p,s)=e^{-ip_\alpha x^\alpha}\sqrt{\frac{E+M}{2m}}\left(
\begin{array}{c}
I\\
\frac{\vec\sigma\cdot\vec{p}}{E+M}
\end{array}
\right)\chi_s.
\end{align}	
Correspondingly we express the non-relativistic Pauli wave function as	
\begin{align}%(7.4)
\Psi=\sum_{s=1,2}\limits\int\frac{d^3p}{(2\pi)^3}f(p,s)e^{ip_\alpha x^\alpha}\chi_s.
\end{align}		
In terms of the above expressions (7.2) and (7.3) the interaction energy (7.1b) is	
\begin{align}%(7.5)
\Delta E_{EM}=\sum_{s,s'=1,2}\limits&\int\frac{d^3p}{(2\pi)^3}\frac{d^3p'}{(2\pi)^3}f^*(p',s')f(p,s)\notag\\
&[e\int d^3x e^{i(p'_\alpha-p_\alpha)x^\alpha}\bar{u}(p',s')\gamma^\mu u(p,s)A_\mu].
\end{align}	
	
The bracket in (7.5) corresponds to scattering of a free Dirac spinor by an external field, which is equivalent to scattering by an infinitely heavy source particle.  It contains all the information about the spin interaction and corresponds to the diagram in fig. 7.1: the particle leaves the wave function blob with 3-momentum $\vec{p}$, scatters from the external field via the QED vertex amplitude into momentum $\vec{p'}$, and then reenters the wave function blob. The electron remains on the mass shell, corresponding to zero energy transfer, which is consistent with a non-relativistic wave function. We denote the 4-momentum transfer by  $q_\mu=p'_\mu-p_\mu$, with $q_0=0$ . The magnitude of the allowed 3-momentum transfer $\vec{q}$ is limited by the width of the function  $f(p,s)$ in momentum space. 	
	\begin{figure}[h]
 \includegraphics[scale=0.3]{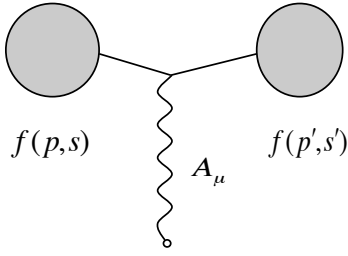}
 \caption{The electron in the wave function scatters from the field and back into the wave function. }
\end{figure}

It is now straightforward to calculate the bracket in (7.5). We split it into 2 parts, $\mu=0$ for the electric interaction and $\mu=j$ for the magnetic interaction. For the electric part we have	
\begin{align}%(7.6)
e\int d^3xA_0 &e^{iq_\alpha x^\alpha}\bar{u}(p',s')\gamma^0u(p,s)\notag\\
=e\int d^3x &e^{iq_\alpha x^\alpha}A_0\left(\frac{E+m}{2m}\right)\chi^\dagger_{s'}\left[I,\frac{\vec\sigma\cdot \vec p'}{E+M}\right]\left[
\begin{array}{c}
I\\\frac{\vec\sigma\cdot \vec p}{E+m}
\end{array}
\right]\chi_s\notag\\
=e\int d^3 x &e^{iq_\alpha x^\alpha}\notag\\
&A_0\chi^\dagger_{s'}\left[\frac{E}{m}+\frac{\vec{q}\cdot\vec{p}}{2m(E+m)}+\frac{i\vec{q}\times\vec{p}\cdot\vec\sigma}{2m(E+m)}\right]\chi_s.
\end{align}	
The first term in the bracket in (7.6) is the obvious charge coupling to the Coulomb field. The second and third terms may be simplified. First, because there is no energy transferred  $\vec{p}^2=\vec{p'}^2$, from which it follows that $\vec{p}\cdot\vec{q}=-\vec{q}^2/2$. Secondly the vector $\vec{q}$ multiplying the exponential may be replaced by $i\nabla$ operating on the exponential, after which integration by parts allows us to replace it with $-i\nabla$ operating on the function $A_0$; that is we may replace $\vec{q}A_0\rightarrow-i\nabla A_0$. Thus the second term vanishes since $\nabla^2A_0=0$ in a charge free region for the Lorentz gauge. What remains is, to order $1/m^2$, 	
\begin{align}%(7.7)
e&\int d^3 x A_0 e^{iq_\alpha x^\alpha}\bar{u}(p',s')\gamma^0 u(p,s)\notag\\
&=\int d^3 e^{iq_\alpha x^\alpha}\left[e(\chi^\dagger_{s'}\chi_s)+\frac{e}{4m^2}\nabla\phi_c\times\vec{p}\cdot(\chi^\dagger_{s'}\vec\sigma\chi_s)\right].
\end{align}	
The second term in (7.7) is clearly a nonlocal fine structure correction, which we mentioned in sec. 4 and which will not concern us further in the present work \cite{Shankar1994}. 	
	
The $\mu=j$ magnetic part of the interaction (7.5) is handled in exactly the same way as the electric part. We have	
\begin{align}%(7.8)
e\int d^3x A_j e^{iq_\alpha x^\alpha}&\bar{u}(p',s')\gamma^ju(p,s)\notag\\
=e\int d^3xe^{iq_\alpha x^\alpha}&A_j\left(\frac{E+m}{2m}\right)\chi^\dagger_{s'}\notag\\
&\left[I,\frac{\vec\sigma\cdot \vec p'}{E+m}\right]\left[
\begin{array}{cc}
0              &\sigma^j \\
\sigma^j  &0
\end{array}
\right]\left[
\begin{array}{c}
I\\
\frac{\vec\sigma\cdot \vec p}{E+m}
\end{array}
\right]\chi_s\notag\\
=\int d^3xe^{iq_\alpha x^\alpha}&A_j\left(\frac{e}{2m}\right)\chi^\dagger_{s'}[\sigma^j\vec\sigma\cdot \vec p+\vec\sigma\cdot \vec p'\sigma^j]\chi_s\notag\\
=-\int d^3 xe^{i\vec q_\alpha x^\alpha}&\left(\frac{e}{2m}\right)\chi^\dagger_{s'}[2\vec{p}\cdot\vec{A}+\vec{q}\cdot\vec{A}+i\vec q\times\vec{A}\cdot\vec\sigma]\chi_s.
\end{align}		
We then replace $\vec{q}\rightarrow-i\nabla$  as discussed above and see that the second term in the bracket vanishes in a gauge with $\nabla\cdot\vec{A}=0$, and we are left with	
\begin{align}%(7.9)
&e\int d^3 x A_j e^{iq_\alpha x^\alpha}\bar{u}(p',s')\gamma^j u(p,s)\notag\\
&=-\int d^3x e^{iq_\alpha x^\alpha}\left(\frac{e}{2m}\right)\chi^\dagger_{s'}[2\vec{p}\cdot\vec{A}+\nabla\times\vec{A}\cdot\vec\sigma]\chi_s\notag\\
&=-\int d^3xe^{iq_\alpha x^\alpha}\left[\frac{e}{m}\vec{p}\cdot\vec{A}(\chi^\dagger_{s'}\chi_s)+\frac{e}{2m}\vec{B}\cdot(\chi^\dagger_{s'}\vec\sigma\chi_s)\right].
\end{align}
	
        Finally we combine (7.7) and (7.9) and substitute into (7.5) to obtain, to order $1/m$,	
\begin{align}%(7.10)
\Delta E_{EM}=&\sum_{s,s'=1,2}\limits\int\frac{d^3p}{(2\pi)^3}\frac{d^3p'}{(2\pi)^3}f^*(p',s')f(p,s)\notag\\
&\int d^3xe^{-i(\vec{p'}-\vec{p})\cdot\vec{x}}\chi^\dagger_{s'}[eA_0-\frac{e}{m}\vec{p}\cdot\vec{A}\frac{e}{2m}\vec{B}\cdot\vec\sigma]\chi_s\notag\\
=&\int d^3 x\Psi^\dagger[eA_0-\frac{e}{m}\vec{p}\cdot\vec{A}-\frac{e}{2m}\vec{B}\cdot\vec\sigma]\Psi.
\end{align}	
This is the same result that we discussed in section 3, so we have thus verified that our present approach reproduces the usual result for the electron g factor, g$_e=2$. 	
	
We now work out the non-relativistic limit of the gravitational interaction in (6.8), following the same procedure as for the electromagnetic interaction; we will not include the product of the electromagnetic and gravitational fields, that is the cross term in (6.8). The algebra is a bit lengthier but equally straightforward. As with the Lagrangian and energy for the electromagnetic case in (7.1) we have for the gravitational case
\begin{align}%(7.11)
L_{IG}=-\frac{1}{2}h_{\mu\nu}T^{\mu\nu},\text{\space\space}\Delta E_G =-\int L_{IG}d^3x,
\end{align}	
where $T^{\mu\nu}$ is given in (6.9). It is convenient to write $T^{\mu\nu}$ in close analogy with the electromagnetic current, as  	
\begin{align}%(7.12)
T^{\mu\alpha}=\bar{\psi}\hat{\gamma}^\alpha(\frac{1}{2}i\overleftrightarrow{\partial}^\mu)\psi.
\end{align}	
Note the relation between the electromagnetic and the gravitational interactions,  	
\begin{align}%(7.13)
A_\mu\leftrightarrow h_{\mu\nu}/2,\text{\space\space}\gamma^\mu\leftrightarrow\gamma^\mu(\frac{i}{2}\overleftrightarrow{\partial}^\nu).
\end{align}	
Then $\Delta E_G$ is, in analogy with (7.5),	
\begin{align}%(7.14)
\Delta E_G&=\sum_{s,s'=1,2}\limits\int\frac{d^3p}{(2\pi)^3}\frac{d^3p'}{(2\pi)^3}f^*(p',s')f(p,s)\notag\\
&[\int d^3x e^{i(p'_\alpha-p_\alpha)x^\alpha}\bar{u}(p',s')\gamma^\mu(p^\nu+q^\nu/2)u(p,s)(h_{\mu\nu}/2)].
\end{align}	
As with the electromagnetism calculation we split the gravitational interaction into two parts, the gravitoelectric for $h_{00}=h_{ii}=2\phi$ and the gravitomagnetic for $h_{0j}=h_{j0}=h^j$. The gravitoelectric part of the bracket in (7.14) involves the same spin products as encountered with the electromagnetic calculation in (7.7) and (7.9), and after some algebra we obtain, to order $1/m^2$,	
\begin{align}%(7.15)
[\int d^3x e^{i(p'_\alpha-p_\alpha)x^\alpha}&\bar{u}(p',s')\{\gamma^0E+(p^j+\frac{q^j}{2})\gamma^j\}u(p,s)\phi]\notag\\
=\int d^3 x e^{i(p'_\alpha-p_\alpha)x^\alpha}&\chi^\dagger_{s'}[\left(\frac{E^2}{m}\phi+\frac{E}{4m^2}\nabla\phi\times \vec p\cdot\vec\sigma\right)\notag\\
&+\left(\frac{\vec{p}^2}{m}\phi+\frac{1}{2m}\nabla\phi\times \vec p\cdot\vec\sigma\right)]\chi_s\notag\\
=\int d^3xe^{i(p'_\alpha-p_\alpha)x^\alpha}&m\chi^\dagger_{s'}[(1+\frac{2\vec{p}^2}{m^2})\phi+\frac{3}{4m^2}\nabla\phi\times \vec p\cdot\vec\sigma]\chi_s.
\end{align}		

A word is in order about the physical interpretation of the gravitoelectric result (7.15). The term $m\phi$ is of course the expected Newtonian energy; the factor $(1+2\vec{p}^2/m^2)$ occurs also in the analysis of a spin zero system in ref. \cite{Adler2010}, and is approximately the Lorentz transformation factor between the potential in the lab frame and the moving frame of the particle; thus $(1+2\vec{p}^2/m^2)\phi$ is the Newtonian potential seen by the moving particle. The last term in the bracket has the same form and is the gravitational analog of the fine structure term in the electromagnetic energy (7.7), except of course for the different coefficient. We will not be concerned further with the higher order terms in (7.15) and will henceforth keep only the lowest order term $\phi$ in the bracket. 

	We turn finally to the gravitomagnetic part of the interaction (7.14), which is our main interest in this work. The gravitomagnetic part of the bracket, proportional to $h^j$, is 
\begin{align}%(7.16)
[\int d^3xe^{i(p'_\alpha-p_\alpha)x^\alpha}\bar{u}(p',s')\{\gamma^0(p^j+q^j/2)&+E\gamma^j\}\notag\\
u(p,s)&(h^j/2)]\notag\\
=[\int d^3x e^{i(p'_\alpha-p_\alpha)x^\alpha}u^\dagger(p',s')\{(p^j+q^j/2)(&h^j/2)\notag\\
+E(h^j/2)\alpha^j&\}u(p,s)].
\end{align}	
Note that the term $\vec{q}\cdot\vec{h}$ will vanish by gauge choice, just as the $\vec{q}\cdot\vec{A}$ term vanished for the electromagnetic case. Then, using the same manipulations as previously on the spin products we reduce this to
\begin{align}%(7.17)
[\int d^3xe^{i(p'_\alpha-p_\alpha)x^\alpha}\bar{u}(p',s')\{\gamma^0(p^j+q^j/2)&+E\gamma^j\}\notag\\
u(p,s)&(h^j/2)]\notag\\
=[\int d^3x e^{i(p'_\alpha-p_\alpha)x^\alpha}\chi^\dagger_{s'}\{\vec{p}\cdot\vec{h}+\frac{1}{4}\nabla\times&\vec{h}\cdot\vec\sigma\}\chi],
	\end{align}
where we have neglected terms of higher order, that is  $1/m^2$. Finally we combine (7.15) and (7.17) to obtain the total energy
\begin{align}%(7.18)
\Delta E_G&=\sum_{s,s'=1,2}\limits\int\frac{d^3p}{(2\pi)^3}\frac{d^3p'}{(2\pi)^3}f^*(p',s')f(p,s)\notag\\
&[\int d^3x e^{i(p'_\alpha-p_\alpha)x^\alpha}\chi^\dagger_{s'}(m\phi+\vec{p}\cdot\vec{h}+\frac{1}{4}\nabla\times\vec{h}\cdot\vec\sigma)\chi]\notag\\
&=\int d^3x \Psi^\dagger(m\phi+\vec{p}\cdot\vec{h}+\frac{1}{4}\vec{\Omega}\cdot\vec\sigma)\Psi.
\end{align}
(Recall that the gravitomagnetic field is $\vec{\Omega}=\nabla\times\vec{h}$.) This is the main result of this section and is consistent with the result of ref. \cite{Adler2010} for a scalar particle.

	Finally we note that since we have expanded the wave function in terms of a free Dirac particle on the mass shell (7.2) the free Dirac Lagrangian is zero and the extra geometric interaction term $(h/2)L$ discussed in section 6 vanishes. 

\section{Gravitomagnetic physical effects}
\setcounter{equation}{0}	

The result (7.18) is to be compared with the analogous electromagnetic result (7.10). We see, of course, that the Newtonian potential is the analog of the Coulomb potential $eA^0$ and the gravitomagnetic potential is the analog of the vector potential according to
\begin{align}%(8.1)
eA^0\leftrightarrow\phi,\text{\space\space}(-e/m)\vec{A}\leftrightarrow\vec{h}.
\end{align}
We also see that the coupling of the spin to the   gravitomagnetic field $\vec{\Omega}$ is only half the analogous electromagnetic coupling. To make this most obvious we consider a gravitomagnetic field $\vec\Omega$ that is approximately constant over the system so that we may choose $\vec{h}=(\vec{\Omega}\times \vec r)/2$. Then 
\begin{align}%(8.2)
\Delta E_G&=\int d^3x\Psi^\dagger(m\phi+\frac{1}{2}\vec{\Omega}\times\vec{r}\cdot\vec{p}+\frac{1}{4}\vec{\Omega}\cdot\vec\sigma)\Psi\notag\\
&=\int d^3x\Psi^\dagger(m\phi+\frac{1}{2}\vec{\Omega}\cdot\vec{r}\times\vec{p}+\frac{1}{2}\vec{\Omega}\cdot\frac{\vec\sigma}{2})\Psi\notag\\
&=\int d^3x\Psi^\dagger[m\phi+\frac{1}{2}\vec{\Omega}\cdot(\vec L+\vec S)]\Psi
\end{align}
Both orbital and spin angular momenta couple in the same way to the gravitomagnetic field, so there is no anomalous g factor for gravitomagnetism; that is g$_\text{g}=1$ for both orbital and spin angular momenta.

From the above correspondence it is clear that since a magnetic moment due to orbital angular momentum, $(e/2m)\vec{L}$, precesses at the Larmor frequency $(eB/2m)$ in a magnetic field $B$, the gravitomagnetic moment due to both orbital and spin angular momenta will precess in a gravitomagnetic field $\Omega$ with frequency $\Omega$, but in the opposite direction. Thus quantum precession should be the same as that observed in the classical gyroscope systems of the GPB satellite experiment \cite{Everitt2011}. It thus seems very likely that the precession rate is universal for any angular momentum system, whether the angular momentum is classical or quantum mechanical, orbital or spin. 

For the surface of the earth the magnitude of the gravitomagnetic field is quite small, as estimated in ref. \cite{Adler2010} The field and the associated quantum energy are of order
\begin{align}%(8.3)
\Omega\approx10^{-13}rad/s,\text{\space\space}E_\Omega=\hbar\Omega\approx10^{-28}eV.
\end{align}
Experimental detection of such small quantum gravitomagnetic effects in an earth-based lab would obviously be difficult. Such an experiment might be performed with an atomic interferometer. The atomic beam could be split into two components with angular momenta differing by $\Delta L=\hbar$. Then, according to (8.2) the two components would have energies differing by about $\Delta E\approx\Omega\Delta L\approx\Omega\hbar$ and thus suffer phase shifts differing by about $\Delta\varphi\approx\Delta Et/\hbar\approx\Omega t$, where $t$  is the time of flight. For a typical $t=1s$ this implies a phase shift of order $10^{-13}rad$, which is orders of magnitude less than presently detectable \cite{Adler2011}. 

In addition to the small size of gravitomagnetic effects one might see in the laboratory there is a serious further inherent difficulty in almost any such experiment; a rotation of the apparatus would in general have similar effects and swamp the gravitomagnetic effects, so such rotations would have to be controlled and compensated to very high accuracy as mentioned in the introduction and in ref. \cite{Adler2010}. 

The results of the GPB experiment and the theoretical results of this paper and ref.\cite{Adler2010} are probably most important in establishing the validity and consistency of general relativity and the gravitomagnetic effects that it implies. Such gavitomagnetic effects are quite small in earth-based labs and satellite systems, as is clear from (8.3), but may play a large role in astrophysical phenomena such as the jets observed in active galactic nuclei, for which the gravitomagnetic fields are much stronger \cite{Throne2009}.  
\section{Summary and conclusions}
\setcounter{equation}{0}	

We have developed the theory of a spin $1/2$ Dirac particle in a Riemann space and its weak field limit in considerable detail. In the low velocity limit for the particle the energies due to the Newtonian or gravitoelectric field and the Òframe-draggingÓ or gravitomagnetic field take simple and intuitive forms. Detection of the small gravitomagnetic effects in earth-based or satellite experiments is quite difficult, but such effects are expected to be large and of great interest in astrophysical systems such as jets from active galactic nuclei and black holes. 
\\
\setcounter{section}{0}	
\renewcommand{\thesection}{Appendix \Alph{section}}
\renewcommand{\theequation}{\Alph{section}.\arabic{equation}}

\section{The inverse differential operator}
We briefly study the type of differential operator that appears in (3.7) by solving the differential equation 
\begin{align}%(A.1)
Af+\partial f=(A+\partial)f=F,\text{\space\space}f=f(x),\text{\space\space}F=F(x),
\end{align}
where $F(x)$ is a given function that may be expanded as a power series in the region of interest and $A$ is a constant. The solution of the homogeneous equation is
\begin{align}%(A.2)
f_h=Ce^{-Ax}\text{\space\space}(C=\text{arbitrary constant}).
\end{align}
The general solution of (A.1) is $f_h$ plus any particular solution $f_p$; for the particular solution we solve (A.1) symbolically as, 
\begin{align}%(A.3)
f_p=(A+\partial)^{-1}F=\frac{1}{A}\left(1-\frac{\partial F}{A}+\frac{\partial^2F}{A^2}...\right).
\end{align}
Operating on (A.3) with $(A+\partial)$ obviously gives $F$. 

	To further justify the above formal operations we may solve (A.1) in a different way. An integrating factor is easily seen to be $e^{Ax}$, so
\begin{align}%(A.4)
\partial(e^{Ax}f)=e^{Ax}(A+\partial)f=e^{Ax}F.
\end{align}
Integration then gives the general solution 
\begin{align}%(A.5)
f=e^{-Ax}\int^x\limits e^{-Ax'}F(x')dx'+Ce^{-Ax}.
\end{align}
Since (A.1) is linear and $F$ is assumed to be expandable in a power series we need only consider powers, $F=x^n$. Then we easily evaluate (A.5) using integration by parts, to obtain
\begin{align}%(A.6)
f=\frac{1}{A}\left(\frac{x^n}{A}-\frac{nx^{n-1}}{A^2}+\frac{n(n-1)x^{n-2}}{A^3}...+1\right)+Ce^{-Ax}.
\end{align}
This agrees with the power series for $f_p$ given in (A.3). 

\section{Energy momentum tensor for the Dirac field}
\setcounter{equation}{0}	

We wish to obtain the energy momentum tensor for a Dirac field in flat space, which occurs in (6.8) and (6.9)\cite{Bjorken1964}. We begin with the Lagrangian (3.1) for the free Dirac field and work out the canonical energy momentum tensor according to the Noether theorem; it is, up to a constant multiplier $C$,
\begin{align}%(B.1)
T^\mu_{\text{\space}\nu}=C[\frac{\partial L}{\partial\psi_{|\mu}}\psi_{|\nu}+\frac{\partial L}{\partial\bar{\psi}_{|\mu}}\bar{\psi}_{|\nu}-\delta^\mu_\nu L]\notag\\
=C[a\bar{\psi}i\gamma^\mu\psi_{|\nu}-b\bar{\psi}_{|\nu}i\gamma^\mu\psi].
\end{align}
where we have omitted the term proportional to $L$ since it is zero for a solution of the free Dirac equation. Using the fact that the Dirac and the Klein-Gordon equations are obeyed by $\psi$ we calculate the two divergences of this tensor to be 
\begin{align}%(B.2)
T^{\mu\nu}_{\space\space\space}|{\mu}=0,\text{\space\space}T^{\mu\nu}_{\space\space\space}|{\nu}=C(b-a)[m^2(\bar{\psi}i\gamma^\mu\psi)-(\bar{\psi}^{|\nu}i\gamma^\mu\psi_{|\nu})]
\end{align}
If we choose $b=a$, as in the text, both divergences are zero and the tensor has symmetry in $\psi$ and $\bar{\psi}$. Moreover we may then consistently symmetrize $T^{\mu\nu}$ and have 
\begin{align}%(B.3)
T^{\mu\nu}=\frac{1}{4}[\bar{\psi}i\gamma^\mu\psi^{|\nu}-\bar{\psi}^{|\nu}i\gamma^\mu\psi+\bar{\psi}i\gamma^\nu\psi^{|\mu}-\bar{\psi}^{|\mu}i\gamma^\nu\psi]
\end{align}
This has now been normalized so that in the low velocity limit 
\begin{align}%(B.4)
T^{00}\approx m\bar{\psi}\psi
\end{align}
Finally, to include the electromagnetic field we use the minimal substitution recipe $i\partial_\mu\rightarrow i\partial_\mu-eA_\mu$ to get 
\begin{align}%(B.4)
T^{\mu\nu}&=\frac{1}{4}[\bar{\psi}i\gamma^\mu\psi^{|\nu}-\bar{\psi}^{|\nu}i\gamma^\mu\psi+\bar{\psi}i\gamma^\nu\psi^{|\mu}-\bar{\psi}^{|\mu}i\gamma^\nu\psi]\notag\\
&-\frac{1}{2}[e\bar{\psi}A^\nu\gamma^\mu\psi+e\bar{\psi}A^\mu\gamma^\nu\psi]
\end{align}
 To verify the result (B.5) we may calculate the divergence of $T^{\mu\nu}$ to find, after some algebra, that it gives the correct Lorentz force, 
\begin{align}%(B.6)
T^{\mu\nu}_{\text{\space\space\space}|\mu}=-j_\alpha F^{\mu\alpha}=-(\bar{\psi}\gamma_\alpha\psi)F^{\mu\alpha}
\end{align}
 In the interaction Lagrangian (6.8) the energy momentum tensor is contracted with the symmetric $h_{\mu\nu}$ so the symmetrization in (B.5) is not relevant. 
  
\section*{Acknowledgements}
This work was partially supported by NASA grant 8-39225 to Gravity Probe B and by NSC of Taiwan under Project No. NSC 97-2112-M-002-026-MY3. Pisin Chen thanks Taiwan's National Center for Theoretical Sciences for their support. Thanks go to Robert Wagoner, Francis Everitt, and Alex Silbergleit and other members of the Gravity Probe B theory group for useful discussions, and to Mark Kasevich of the Stanford physics department for interesting comments on atomic beam interferometry and equivalence principle experiments. Kung-Yi Su provided valuable help with the manuscript. Elisa Varani thanks Cavallo Pacific for encouragement and support.

%\section*{References: (whose format is to be used?)}  


\begin{thebibliography}{99}
%1
\bibitem{Adler2006} R. J. Adler, ``Gravity,'' chapter 2 of {\it The New Physics for the Twenty-first Century}, edited by Gordon Fraser, (Cambridge University Press, Cambridge UK, 2006).
%2
\bibitem{WILL1993}C. Will, {\it Theory and Experiment in Gravitational Physics}, (Cambridge University Press, Cambridge UK, 1981, revised edition 1993), see chapters 2 and 5. 
%3
\bibitem{Ohanian1976} H. C. Ohanian and R. Ruffini, {\it Gravitation and Spacetime}, (W. W. Norton, New York, 1976), chapter 1. 
%4
\bibitem{Everitt2011}C. W. F. Everitt et. al. Phys. Rev. Lett. {\bf 106}, 221101(2011); see also the GPB website: \href{http://einstein.stanford.edu}{http://einstein.stanford.edu}. 
%5
\bibitem{Will43} C. Will, ``Finally, Results from Gravity Probe B", at \href{http//:physics.aps.org/articles/v4/43}{http://physics.aps.org/articles/v4/43}
%6
\bibitem{Ciufolini1997} I. Ciufolini et. al. Class. Quantum Gravit. $\bf 14$, 2701 (1997);
I. Ciufolini et. al. Science $\bf 279$, 2100 (1998); I. Ciufolini and J. A. Wheeler, {\it Gravitation and Inertia}, (Princeton University Press, New Jersey, 1995), chapter 6; I. Ciufolini et. al., in {\it General Relativity and John Archibald Wheeler}, edited I. Ciufolini and R. A. Matzner, (Springer, Dordrecht, 2010), p. 371. 
%7
\bibitem{Iorio2004} L. Iorio, H. I. M. Lichtenegger, and C. Corda, {\it Astrophys. Space Sci.}, {\bf 331}, 351 (2011), discusses the phenomenology of solar system  tests of general relativity; I. Ciufolini and E. C. Pavlis, Nature, {\bf 43}, 958, (2004). 
%8
\bibitem{Throne2009}K. S. Thorne, {\it Near Zero: New Frontiers of Physics}, edited by J. D. Fairbank, B. S. Deaver, C. W. F. Everitt and P. F. Michelson (W. H. Freeman, New York, 1988) p. 573; L. Stella and A. Possenti, {\it Space Sci. Rev.}, 148 (2009); R. D. Blandford and R. L. Znajek, Mo. Not. Roy. Astro. Soc., {\bf 179}, 433 (1977); R. K. Williams, (1995, May 15). Phys. Rev., {\bf 51}(10), 5387-5427, (1995).
%9
\bibitem{Hawking1970} S. W. Hawking, Commun. Math. Phys. {\bf 18}, 301 (1970).
%10
\bibitem{Birell1982} N. D. Birrell and P. C. W. Davies, {\it Quantum Fields in Curved Space}, (Cambridge University Press, Cambridge UK, 1982).
%11
\bibitem{Adler2001} R. J. Adler, P. Chen, and D. Santiago, Gen. Rel. and Grav. {\bf 33}, (2001).
%12
\bibitem{Nesvizhevdky2002} V. Nesvizhevdky et. al. Nature {\bf 415}, 297-299, (2002),
%13
\bibitem{Dimopoulos2008} S. Dimopoulos, P. W. Graham, J. M. Hogan, and M. A. Kasevich,  Phys. Rev. Lett. {\bf 98}, 111102 (2007); S. Dimopoulos, P. W. Graham, J. M. Hogan, M. A. Kasevich, and S. Rajendran, Phys. Rev. D {\bf 78}, 12202 (2008). 
%14
\bibitem{Weinberg1972} S. Weinberg, {\it Gravitation and Cosmology}, (John Wiley, New York, 1972), see chapter 5 on effects of general relativity on the motion of a classical particle. 
%15
\bibitem{Adler2010} R. J. Adler and P. Chen, Phys. Rev. D, {\bf 82} 025004 (2010). 
%16
\bibitem{Bjorken1964} J. D. Bjorken and S. D. Drell, {\it Relativistic Quantum Mechanics}, (McGraw Hill, New York, 1964), we use the conventions and notation for the Dirac equation in flat space given in chapter 3; J. D. Bjorken and S. D. Drell, {\it Relativistic Quantum Fields}, (McGraw Hill, New York, 1965), chapter 13.
%17
\bibitem{Lawrie1990} I. Lawrie, {\it A Unified Grand Tour of Theoretical Physics}, (Adam Hilger, Bristol, 1990), generally covariant Dirac theory using tetrads is discussed in sec. 7.5; see also ref. [14] sec. 12.5. 
%18
\bibitem{Misner1970} C. W. Misner, K. S. Thorne, and J. A. Wheeler, {\it Gravitation}, (W. H. Freeman, San Francisco, 1970); R. J. Adler, M. Bazin, and M. M. Schiffer, {\it Introduction to General Relativity, 2nd edition}, (McGraw Hill, New York, 1975), see chapter 9. 
%19
\bibitem{Adler1999} R. J. Adler, Gen. Rel. and Grav. {\bf 31}, 1837 (1999.) 
%20
\bibitem{Adler2000} R. J. Adler and A. S. Silbergleit Int. J. Th. Phys. {\bf 39}, 1291 (2000). Note that the definition of the gravitomagnetic field differs by a factor of 2 from our use in this work. 
%21
\bibitem{Feynman1965} R. P. Feynman, {\it Lectures on Gravitation}, lecture notes by F. B. Morinigo and W. C. Wagner (Caltech, Pasadena, 1971); See also ref. \cite{Weinberg1972}, sec. 6.9; P. C. Peters, {\it Flat-space Gravitation and Feynman Quantization}, lecture notes, (1965).  
%22
\bibitem{Oriti2009} An overview of this vast field is given in D. Oriti, {\it Approaches to Quantum Gravity}, (Cambridge University Press, Cambridge UK, 2009). 
%23
\bibitem{Trautman2010} A. Trautman, ``Einstein-Cartan Theory,''  in {\it Encyclopedia of Mathematical Physics}, edited by J.-P. Francoise, G. L. Naber, and S. T. Tsou, (Elsevier, Oxford UK, 2006), p. 189; H. Kleinert, [gr-qc] arxiv:1005.1460 (2010)
%24
\bibitem{Peskin1995} M. E. Peskin and D. V. Schroeder, {\it An Introduction to Quantum Field Theory}, (Addison-Wesley, Reading MA, 1995), chapter 3. 
%25
\bibitem{Shankar1994} R. Shankar, {\it Principles of Quantum Mechanics, 2$^{\text{nd}}$ ed}. (Plenum, New York, 1994), section 20.2.
%26
\bibitem{Schwinger1976} J. Schwinger, Am. J. Phys. {\bf 42}, 507 (1973); J. Schwinger, Gen. Rel. and Gravit. 7, 251 (1976). 
%27
\bibitem{Adler1966} R. J. Adler and S. D. Drell, Phys. Rev. Lett. {\bf 13}, 349, (1964); R. J. Adler, Phys. Rev. {\bf 141}, 1499-1508 (1966).
%28
\bibitem{Adler2011} R. J. Adler, H. Mueller, and M. L. Perl, Int. J. of Mod. Phys. A, (2011); to be published.
%29
\bibitem{Thorne16} See also chapter 5 in ref. [16] 

   \end{thebibliography}
\end{document}